\documentclass[pra,tighten,nofootinbib,groupedaddress,amssymb,amsmath]{revtex4-1}

\usepackage{dsfont}
\usepackage{bm}
\usepackage{bbm}
\usepackage[mathscr]{euscript}
\usepackage{amsmath}
\usepackage{amssymb}
\usepackage{amsthm}
\usepackage{amsfonts}
\usepackage{graphicx}
\usepackage{color}
\usepackage[colorlinks=true,urlcolor=blue,linkcolor=red,citecolor=green]{hyperref}
\begin{document}
\title{Hedging and Leveraging:  Principal Portfolios of the Capital Asset Pricing Model}
\author{M. Hossein Partovi}
\affiliation{Department of Physics and Astronomy, California State
University, Sacramento, California 95819-6041}

\begin{abstract}

The principal portfolios of the standard Capital Asset Pricing Model
(CAPM) are analyzed and found to have remarkable hedging and
leveraging properties. Principal portfolios implement a recasting of any \textit{correlated} asset set of $N$
risky securities into an equivalent but \textit{uncorrelated} set
when short sales are allowed.  While a determination of principal
portfolios in general requires a detailed knowledge of the
covariance matrix for the asset set, the rather simple structure of
CAPM permits an accurate solution for any reasonably large asset set that reveals
interesting universal properties.  Thus for an asset set of size $N$, we find a
\textit{market-aligned} portfolio, corresponding to the \textit{market}
portfolio of CAPM, as well $N-1$ \textit{market-orthogonal}
portfolios which are market neutral and strongly leveraged. These
results provide new insight into the return-volatility structure of
CAPM, and demonstrate the effect of unbridled leveraging on
volatility.

\end{abstract}
\maketitle

\section*{1. Introduction}

Modern investment theory dates back to the mean-variance analysis of
Markowitz (1952, 1959), which is expected to hold if asset prices are normally
distributed or the investor preferences are quadratic. Undoubtedly,
the most consequential fruit of Markowitz' seminal work was the
introduction of the capital asset pricing model (CAPM) by Sharpe
(1964), Lintner (1965), and Mossin (1966).  The key ideas of this
model are that investors are mean-variance optimizers facing a
frictionless market with full agreement on the distribution of
security returns and unrestricted access to borrowing and lending at
the riskless rate.  As an asset pricing model, CAPM is an
equilibrium model valid for a given investment horizon, which is
taken to be the same for all investors.  Indeed investors are solely
distinguished by their level of risk aversion.

Principal portfolio analysis, on the other hand, simplifies asset allocation by recasting the asset set into uncorrelated portfolios when short sales are allowed (Partovi and Caputo 2004).   Stated otherwise, the original problem of stock selection from a set of \textit{correlated assets} is transformed into the much simpler problem of choosing from a set of \textit{uncorrelated portfolios}.  The details of this transformation are given in Partovi and Caputo (2004), where the results are summarized as follows:   \textit{Every investment environment ${ \{ {s}_{i}, {r}_{i}, {\sf \sigma}_{ij} \} }_{i,j=1}^{N}$ which allows short sales can be recast as a principal portfolio environment ${ \{ {S}_{\mu}, {R}_{\mu}, {\sf V}_{\mu \nu} \} }_{\mu, \nu =1}^{N}$ where the principal covariance matrix ${\sf V}$ is diagonal.  The weighted mean of the principal variances equals the mean variance of the original environment.  In general, a typical principal portfolio is hedged and leveraged.}   Here $s_i$ ($ {S}_{\mu}$), $r_i$ (${R}_{\mu}$), and ${\sigma}_{ij}$ (${\sf V}_{\mu \nu} $) represent the assets, the expected returns, and the covariance matrix of the original (recast) set, while $N$ is the size of the asset set.   It was further shown in Partovi and Caputo (2004) that the efficient frontier in the presence of a riskless asset has a simple allocation rule which requires that each principal portfolio be included in inverse proportion to its variance. Practical applications of principal portfolios have already been considered by several authors, for example, Poddig and Unger (2012) and Kind (2013).

In this paper we present a perturbative calculation of the principal portfolios of the single-index CAPM in the large $N$ limit.  The results of this calculation are in general expected to entail a relative error of the order of $1/{N}^2$.   However, since any application of the single-index CAPM is most likely to involve a large asset set, the stated error is normally quite small and in any case majorized by modelling errors.   Thus the results to be reported here are accurate implications of the underlying model.

The principal portfolio analysis of the single-index model and an exactly solvable version of it presented in \S 3 highlight the volatility structure of principal portfolios in a practical and familiar context.  A
remarkable result of the analysis is the bifurcation of the set of principal portfolios into a {\it market-aligned} portfolio, which is
unleveraged and behaves rather like a total-market index fund, and
$N-1$ \textit{market-orthogonal} portfolios, which are hedged and
leveraged,\footnote{We use the term
``leveraged'' here to imply that the portfolio contains borrowed
assets, e.g., short-sold positions.} and nearly free of market driven fluctuations. This
equivalency between the original asset set and two classes of
principal portfolios is reminiscent of, but fundamentally different
from, Merton's (1972) two mutual fund theorems.  The
market-orthogonal portfolios, on the other hand, provide a vivid demonstration of the
effect of leveraging on the volatility level of a portfolio.

\section*{2. Principal Portfolios of the Single-Index Model}

Here we shall analyze the standard single-index model as well as an exactly solvable
special case of it with respect to their principal portfolio structure.
Remarkably, our analysis will uncover interesting and hitherto unnoticed
properties of well-diversified and arbitrarily leveraged portfolios within
the single-index model.

Consider a set of $N$ assets $ \{ {s}_{i} \}$, $1 \le i \le N$, whose
rates of return are normally distributed random variables given by
\begin{equation}
{\rho}_{i}\stackrel{\rm def}{=}{\alpha}_{i}+{\beta}_{i}{\rho}_{mkt},
\label{431}
\end{equation}
where ${\alpha}_{i}$ and ${\rho}_{mkt}$ are uncorrelated, normally
distributed
random variables with expected values and variances equal to
${\bar{\alpha}}_{i}$, ${\bar{\rho}}_{mkt}$ and
${\bar{{\alpha}_{i}^{2}}}$,
${\bar{{\rho}^{2}}}_{mkt}$, respectively.  The quantity ${\beta}_{i}$
associated with asset ${s}_{i}$ is a constant which measures the degree
to which ${s}_{i}$ is coupled to the overall market variations.  Thus the
attributes of a given asset are assumed to consist of a {\it market-driven} (or {\it systematic}) part described by
$({\beta}_{i}{\rho}_{mkt},{\beta}_{i}^{2}{\bar{{\rho}^{2}}}_{mkt})$ and a
{\it residual} (or {\it specific}) part described by
$({\alpha}_{i},{\bar{{\alpha}_{i}^{2}}})$, with the two parts being
uncorrelated.

The expected value of Eq.~(\ref{431}) is given by
\begin{equation}
{\bar{{\rho}_{i}}}\stackrel{\rm
def}{=}{r}_{i}={\bar{{\alpha}_{i}}}+{\beta}_{i}{\bar{\rho}}_{mkt}.
\label{4311}
\end{equation}
The covariance matrix which results from Eq.~(\ref{431}) is
similarly a superposition of the specific and market-driven
contributions, as would be expected of the sum of two uncorrelated
variables.  It can be written as
\begin{equation}
{\sf \sigma}_{ij}={\bar{{\alpha}_{i}^{2}}} {\delta}_{ij}+
{\beta}_{i}{\beta}_{j} {\bar{{\rho}^{2}}}_{mkt}.  \label{432}
\end{equation}
Note that ${\sf \sigma}$ is a {\it definite} matrix, since we have
excluded riskless assets
from the asset set for the time being.

We shall assume here that the number of assets $N$ is appropriately
large,
as is in fact implicit in the formulation of all index models, so that
the
condition ${\bar{{\alpha}_{i}^{2}}}/ N {b}^{2} {\bar{{\rho}^{2}}}_{mkt}
\ll 1$ is
satisfied; here $b\stackrel{\rm def}{=}
{({\sum}_{i=1}^{N}{\beta}_{i}^{2}/N)}^{1 \over 2}$ is the square root of
the
average value of ${\beta}_{i}^{2}$, typically of the order of unity.
These assumptions are not essential
to our discussion, but they do simplify the presentation and more
importantly, they are usually well satisfied for appropriately large
values of $N$ and guarantee that our purturbative results below are accurate for practical applications.

Under the above assumptions it is appropriate to rescale
the covariance matrix as in ${\sf \sigma}_{ij}= N {b}^{2}
{\bar{{\rho}^{2}}}_{mkt}
{\tilde{\sf \sigma}}_{ij}$, where
\begin{equation}
{\tilde{\sf \sigma}}_{ij}\stackrel{\rm def}{=}{\gamma}_{i}^{2}
{\delta}_{ij}+
{\hat{\beta}}_{i}{\hat{\beta}}_{j}  \label{433}
\end{equation}
is a dimensionless matrix.  Here ${\hat{\beta}}_{i} \stackrel{\rm
def}{=}{{\beta}}_{i}/ {({\sum}_{i=1}^{N}{\beta}_{i}^{2})}^{1 \over
2}$, so that $\hat{{\bm { \beta}}}=({\hat{\beta}}_{1},{\hat{\beta}}_{2},
\ldots, {\hat{\beta}}_{N})$ is a unit vector, and
${\gamma}_{i}^{2}\stackrel{\rm def}{=}{\bar{{\alpha}_{i}^{2}}}/ N
{b}^{2} {\bar{{\rho}^{2}}}_{mkt} \ll 1$ as concluded above.

The above representation of the covariance matrix for the single-index
model is quite suitable for revealing the structure of its eigenvalues
and
eigenvectors, these being closely related to the rescaled principal
variances
and the principal portfolios we are seeking.  The structure in question
is
actually discernible on the basis of simple, qualitative considerations
of
the spectrum of ${\tilde{\sf \sigma}}$.  To see this structure, let us
first note that the sum of the eigenvalues of ${\tilde{\sf \sigma}}$,
which is given by ${\rm tr} ({\tilde{\sf \sigma}})\stackrel{\rm def}{=}
{\sum}_{i=1}^{N}{\tilde{\sf \sigma}}_{ii}$, equals $1+{\sum}_{i=1}^{N}
{\gamma}_{i}^{2}$.  We will show below that the largest eigenvalue of
${\tilde{\sf \sigma}}$ is approximately equal to unity, so that the
remaining
$N-1$ eigenvalues have an average value approximately equal to the average of
$\{ {\gamma}_{i}^{2} \}$, which was shown above to be much smaller than
unity as a consequence of the large $N$ assumption.  Thus, barring a
strongly skewed distribution of the latter, which
is all but ruled out for any of the customary asset classes, we find that
the spectrum of ${\tilde{\sf \sigma}}$ consists of a ``major'' eigenvalue
close to unity, and $N-1$ ``minor'' eigenvalues each much smaller than
unity.
Stated in terms of the spectrum of ${\sf V}$, this implies that the
principal portfolios separate into two classes of quite different
properties, namely (i) a single {\it market-aligned} portfolio with a
variance of magnitude approximately equal to
$N{b}^{2}{\bar{{\rho}^{2}}}_{mkt}/{{W}_{N}}^{2}$, and (ii) $N-1$ {\it
market-orthogonal} portfolios whose variances have a weighted average approximately equal
to the average of the residual variance of the original asset set.  As one
might suspect, these two categories are characterized by sharply different
values of portfolio beta,\footnote{Here the portfolio beta is defined to be the weighted mean of beta in the single-index model literature} the former
with a value typical of the asset set (i.e., of the order of unity) and
the remaining $N-1$ portfolios with much smaller (possibly vanishing; cf.
\S 3) values.

To see the quantitative details of the foregoing qualitative analysis, we
now turn to a perturbative treatment of the spectrum of
${\tilde{\sf \sigma}}$.  The eigenvalue equation for ${\tilde{\sf
\sigma}}$ reads
\begin{equation}
{({\tilde{\sf \sigma}}{\mathbf{e}}^{\mu})}_{i}={\gamma}_{i}^{2}
{e}^{\mu}_{i}+ {\hat{\bm{\beta}}}\cdot {\mathbf{{e}^{\mu}}}
{\hat{\beta}}_{i} ={\tilde{v}}_{\mu}^{2}{e}^{\mu}_{i}, \label{434}
\end{equation}
where ${\bm{e}}^{\mu}$ is the $\mu$th eigenvector, ${e}^{\mu}_{i}$
is the $i$th component of that eigenvector, and
${\tilde{v}}_{\mu}^{2}$ is the corresponding eigenvalue, all
quantities as defined earlier.  Because of its simple structure,
Eq.~(\ref{434}) can be implicitly solved for the components of the
eigenvectors to yield
\begin{equation}
{e}^{\mu}_{i}=[{\hat{\bm{\beta}}}\cdot {\mathbf{{e}^{\mu}}} /
({\tilde{v}}_{\mu}^{2}-{\gamma}_{i}^{2})]{\hat{\beta}}_{i}.
\label{435}
\end{equation}
Upon multiplying this equation by ${\hat{\beta}}_{i}$ and summing over
$i$, we find the characteristic equation for the eigenvalues.  It reads
\begin{equation}
1={\sum}_{i=1}^{N}[{\hat{\beta}}_{i}^{2} / ({\tilde{v}}_{\mu}^{2}-
{\gamma}_{i}^{2})].  \label{436}
\end{equation}
This equation can be rearranged as an $N$th-order polynomial
equation in the variable ${\tilde{v}}_{\mu}^{2}$, the $\mu$th
eigenvalue of ${{\sf \sigma}}$ divided by
$N{b}^{2}{\bar{{\rho}^{2}}}_{mkt}$, and is guaranteed to have $N$
real, positive roots (with multiple roots counted according to their
multiplicity).  Once these roots are determined, they can be used in
Eq.~(\ref{435}) to find the eigenvectors in the usual manner.

As mentioned earlier, the structure of ${\tilde{\sf \sigma}}$ allows
an approximate determination of its largest eigenvalue when $N$ is
suitably large, say a hundred or more.  This is of course a
significant advantage in any numerical solution of the equations
described in the preceding paragraph. As one can see from
Eq.~(\ref{433}), the matrix in question, ${\tilde{\sf \sigma}}$, is
the sum of two parts, one is diagonal with elements
${\gamma}_{i}^{2}$ which are much smaller than unity, and the other
a rank-1 matrix with eigenvalue equal to unity.  This implies that
the eigenvector of the latter matrix is an approximate eigenvector
of ${\tilde{\sf \sigma}}$ with eigenvalue approximately equal to
unity.  This is the eigenvalue we designated as {\it major} in our
qualitative discussion. Let this be the $N$th eigenvalue, so that
${\tilde{v}}_{N}^{2}\stackrel{\rm def}{=}1+{\epsilon}_{N}$, with
$\vert {\epsilon}_{N} \vert \ll 1$.  Substituting this expression
for ${\tilde{v}}_{N}^{2}$ in Eq.~(\ref{436}), and treating the
resulting equation to first order in
${\gamma}_{i}^{2}$,\footnote{This is the approximation in which any
contribution to ${\tilde{v}}_{N}^{2}$ whose ratio to
${\gamma}_{i}^{2}$ vanishes in the $N \rightarrow \infty$ limit will
be neglected.} we find
\begin{equation}
{\tilde{v}}_{N}^{2} \simeq 1+{\sum}_{i=1}^{N}
{\gamma}_{i}^{2}{\hat{\beta}}_{i}^{2},  \label{437}
\end{equation}
which identifies ${\epsilon}_{N} $ as equal to ${\sum}_{i=1}^{N}
{\gamma}_{i}^{2}{\hat{\beta}}_{i}^{2}$ to first order, thus
verifying the condition $\vert {\epsilon}_{N} \vert \ll 1$.  The
corresponding eigenvector can now be found from Eq.~(\ref{435}); to
first order, it is given by
\begin{equation}
{e}^{N}_{i} \simeq (1+ {\gamma}_{i}^{2}-{\sum}_{j=1}^{N}
{\gamma}_{j}^{2}{\hat{\beta}}_{j}^{2}){\hat{\beta}}_{i},  \label{438}
\end{equation}
where the conditions of unit length and non-negative relative weight
stipulated earlier have already been imposed within the stated
order of approximation.

Equation (\ref{438}) specifies the (relative) composition of the
market-aligned portfolio.  The relative weight ${W}_{N}$ of this
portfolio, on the other hand, is expected to be of the order of
${N}^{1 \over 2}$, since this portfolio consists entirely of
purchased assets (recall our estimate of the relative weights
earlier in \S 2).  Indeed one can see from Eq.~(\ref{438}) that
${W}_{N} \simeq {\sum}_{i=1}^{N}{\hat{\beta}}_{i}$ in the leading
order of approximation,\footnote{This is the approximation in which
any contribution to ${W}_{N}$ whose ratio to ${\hat{\beta}}_{i}$
vanishes in the $N \rightarrow \infty$ limit will be neglected.}
which confirms the above-stated estimate (recall that the average of
the ${\hat{\beta}}_{i}^{2}$ equals ${N}^{-1}$).  Equations
(\ref{437}) and (\ref{438}) provide approximate expressions for the
major eigenvalue and eigenvector of the covariance matrix of the
single-index model.

Rescaling Eqs.\ (\ref{437}) and (\ref{438}) back to original variables,
we
find, for the variance and the composition of the market-aligned
principal portfolio, the expressions
\begin{equation}
{{V}_{N}}^{2} \simeq [1+3 {\sum}_{i=1}^{N}
{\gamma}_{i}^{2}{\hat{\beta}}_{i}^{2}-{({\sum}_{i=1}^{N}
{\hat{\beta}}_{i})}^{-1}{\sum}_{i=1}^{N}
{\gamma}_{i}^{2}{\hat{\beta}}_{i}]{({\bm{\beta}} \cdot
{\bm{\beta}})} {\bar{{\rho}^{2}}}_{mkt}
/{({\sum}_{i=1}^{N}{\hat{\beta}}_{i})}^{2},  \label{439}
\end{equation}
\begin{equation}
{e}^{N}_{i}/{W}_{N} \simeq [1+
{\gamma}_{i}^{2}-{({\sum}_{i=1}^{N}
{\hat{\beta}}_{i})}^{-1}{\sum}_{i=1}^{N}
{\gamma}_{i}^{2}{\hat{\beta}}_{i}]{\hat{\beta}}_{i}/({\sum}_{j=1}^{N}
{\hat{\beta}}_{j}),  \label{440}
\end{equation}
where we have left the small correction terms in dimensionless form.
It is clear from Eq.~(\ref{440}) that the market-aligned portfolio
is basically composed by investing in each asset in proportion to
how strongly it is correlated with the overall market fluctuations,
i.e., in proportion to the value of its beta; cf. Eq.~(\ref{431}).
Consequently, it is expected to be strongly suseptible to market-driven
fluctuations.  Indeed as one can see from Eq.~(\ref{439}), the
variance of this principal portfolio in the leading order is given
by $(N{b}^{2}/{{W}_{N}}^{2}){\bar{{\rho}^{2}}}_{mkt}$, which is of
the same order of magnitude as ${\bar{{\rho}^{2}}}_{mkt}$ (recall
that $b$ is of the same approximate magnitude as a typical $\beta$
and that ${W}_{N}$ is of the order of ${N}^{1 \over 2}$).  The
market-aligned portfolio is therefore seen to be that principal
portfolio which approximately reflects the volatility profile of the
market as a whole.  Moreover, since it entirely composed of
purchased assets, it is neither hedged nor leveraged.

By contrast, the remaining $N-1$ market-orthogonal principal
portfolios are in general hedged and leveraged, and they are quite
immune to overall market fluctuations\footnote{These
market-orthogonal portfolios essentially eliminate what is
referred to as ``market risk'' in the single-index model jargon}.
In fact, since ${\sum}_{\mu=1}^{N} {{v}}_{\mu}^{2}={\rm tr}({{\sf
\sigma}})={\bm{\beta}} \cdot {\bm{\beta}}
{\bar{{\rho}^{2}}}_{mkt}(1+{\sum}_{i=1}^{N} {\gamma}_{i}^{2})$,
and ${{v}}_{N}^{2} \simeq {\bm{\beta}} \cdot {\bm{\beta}}
{\bar{{\rho}^{2}}}_{mkt} + {\sum}_{i=1}^{N} {\hat{\beta}}_{i}^{2}
{\bar{{\alpha}_{i}^{2}}}$, we find for the the average value of
the $N-1$ minor eigenvalues
\begin{equation}
{(N-1)}^{-1}{\sum}_{\mu=1}^{N-1} {{v}}_{\mu}^{2}={(N-1)}^{-
1}{\sum}_{\mu=1}^{N-1} {{W}_{\mu}}^{2}{{V}_{\mu}}^{2}
\simeq {(N-1)}^{-1}{\sum}_{i=1}^{N}(1-
{\hat{\beta}}_{i}^{2}) {\bar{{\alpha}_{i}^{2}}}.  \label{441}
\end{equation}
Thus the weighted average of principal variances for market-orthogonal
portfolios is approximately equal to (and in fact less than) the average
of the residual variances of the original asset set.  Therefore, these
$N-1$ market-orthogonal principal portfolios are free not only of mutual
correlations with other portfolios but in general also of the volatility
induced by overall market fluctuations.  This feat is possible in part
because of the very special structure of the single-index model which
makes it possible to isolate essentially all of the systematic market
volatility in one portfolio, leaving the remaining $N-1$ portfolios almost
totally immune to systematic market fluctuations.

There is an important caveat with respect to the foregoing statement.
Recall that there is an inverse relationship between ${{V}}_{\mu}$,
defined as the positive square root of ${{V}}_{\mu}^{2}$,
and ${W}_{\mu}$, so that for highly leveraged portfolios which are
characterized by the condition ${W}_{\mu} \ll 1$, the above argument
would imply a principal variance far exceeding the original ones.  Of
course the condition ${W}_{\mu} \ll 1$ that implies such large variances
also implies large expected returns, so that a more sensible comparative
measure under such conditions
is ${\check{V}}_{\mu}\stackrel{\rm
def}{=}{V}_{\mu}/{R}_{\mu}={v}_{\mu}/{\sum}_{i=1}^{N}{e}^{\mu}_
{i}{r}_{i}$, which may be called {\it return-adjusted volatility} of the
principal portfolio.  As expected, the relative weight ${W}_{\mu}$ is no
longer present in this adjusted version of the volatility.

The return-adjusted volatility for the market-aligned portfolio, on the
other hand, can be calculated from Eqs.\ (\ref{431}), (\ref{439}), and
(\ref{440}).  It is given by
\begin{equation}
{\check{V}}_{N} \simeq \{ 1- [{({\bar{{\rho}^{2}}}_{mkt})}^{1 \over
2}/{\bar{\rho}}_{mkt}]{\sum}_{i=1}^{N}
{\gamma}_{i}{\hat{\beta}}_{i} \} {({\bar{{\rho}^{2}}}_{mkt})}^{1 \over
2}/{\bar{\rho}}_{mkt},  \label{442}
\end{equation}
which is approximately equal to ${({\bar{{\rho}^{2}}}_{mkt})}^{1 \over
2}/{\bar{\rho}}_{mkt}$.  This ratio is of course precisely what one would
expect for the approximate value of the return-adjusted volatility of a
portfolio which is aligned with the overall market price movements.

It is appropriate at this point to summarize the properties of the
principal portfolios for the single-index model.

{\bf Proposition 1.} {\it The principal portfolios of the single-index
model
consist of a market-aligned portfolio, which is unleveraged and has a
return-adjusted volatility $\simeq {({\bar{{\rho}^{2}}}_{mkt})}^{1 \over
2}/{\bar{\rho}}_{mkt}$ characteristic of market-driven price movements,
and $N-1$ market-orthogonal portfolios which are hedged and leveraged,
and nearly free of systematic market fluctuations.  Equations
(\ref{439})-(\ref{442}) provide approximate expressions valid to first
order in $1/N$ for the properties of these portfolios.}

\section*{3. Single-Index Model with Constant Residual Variance}

To provide an explicit illustration of the principal portfolio
structure within the single-index model described in the preceding
section, we now turn to an exactly solvable, albeit
oversimplified, version of that model. This model is defined by
the assumption that the residual variance of the $i$th asset in
the original set, ${\bar{{\alpha}_{i}^{2}}}$, is equal to
${\bar{{\alpha}^{2}}}$ for all assets.  Observe that this
assumption does not affect the expected rate of return for the
$i$th asset, which is given by
${r}_{i}={\bar{{\alpha}_{i}}}+{\beta}_{i} {\bar{\rho}}_{mkt}$ as
before.  This simplification will allow us to derive an exact
solution for the model and illustrate the concepts and methods of
the previous section in more explicit terms.  The price for this
simplification is of course the unrealistic assumption of constant
residual variance which defines the model.

The covariance matrix with the above simplification appears as
\begin{equation}
{\sf \sigma}_{ij}^{crv}={\bar{{\alpha}^{2}}} {\delta}_{ij}+
{\beta}_{i}{\beta}_{j} {\bar{{\rho}^{2}}}_{mkt},  \label{443}
\end{equation}
whose rescaled version is
\begin{equation}
{\tilde{\sf \sigma}}_{ij}^{crv}={\gamma}^{2} {\delta}_{ij}+
{\hat{\beta}}_{i}{\hat{\beta}}_{j}  \label{444}
\end{equation}
These equations are of course specialized versions of Eqs.\ (\ref{432})
and (\ref{433}).

Referring to the results of the previous section, one can readily
see that the spectrum of ${\tilde{\sf \sigma}}^{crv}$ consists of a
major eigenvalue (exactly) equal to $1+{\gamma}^{2}$ [cf.
Eq.~(\ref{437})], and $N-1$ minor eigenvalues, all equal to
${\gamma}^{2}$.  Recall that these eigenvalues respectively
correspond to the market-aligned and market-orthogonal portfolios
introduced in \S 4.1.  Not surprisingly, the spectrum of
${\tilde{\sf \sigma}}_{ij}^{crv}$ is found to be highly degenerate.
The eigenvector ${{\bf e}^{crv}}^{N}$ corresponding to the major
eigenvalue is (exactly) equal to ${\hat{\beta}}_{i}$ [cf.
Eq.~(\ref{438})], while the remaining $N-1$ minor eigenvectors are
not uniquely determined\footnote{This is of course the exceptional
case of spectrum degeneracy mentioned in \S 2.} and may be
arbitrarily chosen to be any orthonormal set of $N-1$ vectors
orthogonal to the major eigenvector ${\hat{\beta}}_{i}$.  The
expected return and volatility features of the $N-1$
market-orthogonal portfolios defined by this arbitrary choice, on
the other hand, do depend on that choice, as the following analysis
will show.

Since our main objective is the determination of the efficient
frontier, we shall choose the remaining $N-1$ eigenvectors with
respect to their volatility level, which, it may be recalled from
\S 2, is given by
${{V}_{\mu}}^{2}={v}_{\mu}^{2}/{{W}_{\mu}}^{2}$.  For the present
case, minimizing ${V}_{\mu}$ amounts to maximizing ${W}_{\mu}$.
Therefore, we will look for a unit vector $\bf e$ that is
orthogonal to $\bm{{\hat{\beta}}}$ as stipulated above {\it and}
maximizes ${\sum}_{i=1}^{N}{e}_{i}$.  In terms of rescaled
quantities, this problem appears as
\begin{equation}
{\max \;}_{\bf e} {\;} {\bf e} \cdot {\hat{{\bf u}}} {\:} {\;}
s.t. \; {{\bf e} \cdot {\bf e}}=1, {\:} {\:} {\bf e} \cdot
{\hat{\bm{\beta}}}=0, \label{445}
\end{equation}
where ${\hat{{\bf u}}}_{i}\stackrel{\rm def}{=}{N}^{-{1 \over
2}}(1,1, \ldots,1)$ is an $N$-dimensional unit vector all of whose
components are equal.  The solution to Eq.~(\ref{445}) may be found
by standard methods provided that ${\hat{{\bf u}}}$ and
${\hat{\bm{\beta}}}$ are not parallel, a condition whose violation
is very improbable and will henceforth be assumed to hold.  On the
other hand, it is clear from geometric considerations that the
solution must be that linear combination of ${\hat{{\bf u}}}$ and
${\hat{\bm{\beta}}}$ which is orthogonal to ${\hat{\bm{\beta}}}$.
Designating the solution vector as ${{\bf e}^{crv}}^{1}$, we find
\begin{equation}
{{\bf e}^{crv}_{i}}^{1}= [{\hat{u}}_{i}-\cos (\theta)
{\hat{{\beta}}}_{i}]/ \sin(\theta), \label{446}
\end{equation}
where $\theta$ is the angle formed by the unit vectors ${\hat{{\bf
u}}}$ and ${\hat{\bm{\beta}}}$, constrained by the condition $0 <
\theta \leq \pi /2 $ under our assumptions.  Indeed a little
algebra shows that
\begin{equation}
\tan(\theta) = \delta \beta / {\bar {\beta}}, \label{447}
\end{equation}
where ${\bar{\beta}}$ and $\delta \beta$ respectively denote the
mean and the standard deviation of the $\beta$'s, i.e.,
${N}^{-1}{\sum}_{i=1}^{N} {\beta}_{i}$ and
${[{N}^{-1}{\sum}_{i=1}^{N} {({\beta}_{i}-{\bar {\beta}})}^{2}]}^{1
\over 2}$.  Equation (\ref{447}) clearly shows that the angle
$\theta$ represents the degree of scatter among the betas, vanishing
when all betas are equal and increasing as they are made more
unequal.  Note that the condition $\theta >0$ stipulated above
excludes the (improbable) case of uniform betas. Note also that the
condition ${\bf e} \cdot {\hat{\bm{\beta}}}=0$ in Eq.~(\ref{445}) is
equivalent to the vanishing of the portfolio beta for ${{\bf
e}^{crv}_{i}}^{1}$, in contrast to the same quantity for the
market-aligned portfolio which is found to be ${\bar
{\beta}}/{\cos}^{2}(\theta)$.

Thus far we have determined two eigenvectors, the market-aligned ${{\bf
e}^{crv}}^{N}$, and the minimum-volatility, market-orthogonal eigenvector
${{\bf e}^{crv}}^{1}$.  The remaining $N-2$ market-orthogonal eigenvectors
will of course have to be orthogonal to these, which immediately implies
that they will all be orthogonal to ${\hat{{\bf u}}}$.  But orthogonality
to ${\hat{{\bf u}}}$ implies a vanishing weight, thus implying that these
portfolios require zero initial investment.  Stated differently, these $N-
2$ portfolios are \textit{critically} leveraged, with the short-sold assets precisely
balancing the purchased ones in each portfolio.  Under these
circumstances, any volatility in portfolio return would imply an infinite
variance for that principal portfolio because of the vanishing initial
investment.  Note that the return-adjusted volatility for these
portfolios, on the other hand, need not (and in typical cases will not)
diverge at all.

As stated earlier, the efficient frontier in the presence of a riskless asset has a simple allocation rule which requires that each principal portfolio be
included in inverse proportion to its variance.  For the current
case, this rule clearly excludes the $N-2$ portfolios described
above from the efficient frontier, leaving the first two principal
portfolios and the riskless asset as the only constituents.  Thus
for the special case of constant residual variance, a knowledge of
the two distinguished principal portfolios determined above is all
that is needed to specify the efficient frontier when a riskless
asset is present.  For this reason, we will not continue with the
explicit construction of the remaining $N-2$ eigenvectors.

At this point we can determine the expected value and the variance of the
two principal portfolios determined above according to the definitions and
formulae given in \S 2.  Straightforward algebra leads to
\begin{equation}
{R}^{crv}_{N}={{\sum}_{i=1}^{N}{\hat{{\beta}}}_{i}({\bar{{\alpha}_{i}}}
+{\beta}_{i}{\bar{\rho}}_{mkt}) \over {N}^{{1 \over 2}} \cos
(\theta)}, \;\; {({V}_{N}^{crv})}^{2}={ {\bar{{\alpha}^{2}}} +
{\bm{\beta}} \cdot {\bm{\beta}} {\bar{{\rho}^{2}}}_{mkt} \over N
{\cos}^{2} (\theta)}, \label{448}
\end{equation}
for the market-aligned portfolio, and
\begin{equation}
{R}^{crv}_{1}={{r}^{av} \over {\sin}^{2} (\theta)} -{\cot}^{2}
(\theta){R}^{crv}_{N}
, \;\; {({V}_{1}^{crv})}^{2}={ {\bar{{\alpha}^{2}}} \over N {\sin}^{2}
(\theta)}, \label{449}
\end{equation}
for the market-orthogonal, minimum volatility principal portfolio.  In
order to facilitate comparison with the perturbative results of \S 2 for
the general single-index model, we also record here the
return-adjusted volatilities of these portfolios;
\begin{equation}
{\check{V}}^{crv}_{N} ={{({\bar{{\alpha}^{2}}} + {\bm{\beta}}
\cdot {\bm{\beta}} {\bar{{\rho}^{2}}}_{mkt})}^{1 \over 2} \over
{\sum}_{i=1}^{N}{\hat{{\beta}}}_{i}
({\bar{{\alpha}_{i}}}+{\beta}_{i}{\bar{\rho}}_{mkt})}, \label{450}
\end{equation}
\begin{equation}
{\check{V}}^{crv}_{1} = {{[{\bar{{\alpha}^{2}}} {\sin}^{2} (\theta)]}^{1
\over 2} \over {N}^{-{1 \over 2}}[{r}^{av} -{\cos}^{2}
(\theta){R}^{crv}_{N}] }.
\label{451}
\end{equation}

The results just derived demonstrate the powerful volatility reduction effect of diversification coupled with short sales for the market-orthogonal portfolio ${{\bf e}^{crv}}^{1}$.  To see this,
let us assume a typical value for $\tan (\theta)$ of the order of
unity [for reasonably large $N$; cf. Eq.~(\ref{447})].  We then find
from the above results
\begin{equation}
{\lim}_{N \rightarrow \infty} \; {{V}^{crv}_{N}}={[{{\bm{\beta}}
\cdot {\bm{\beta}} \over N {\cos}^{2} (\theta)}
{\bar{{\rho}^{2}}}_{mkt}]}^{1 \over 2}, \; {\lim}_{N \rightarrow
\infty}\;{{V}^{crv}_{1}}=0. \label{452}
\end{equation}
Note that the quantity ${\bm{\beta}} \cdot {\bm{\beta}}$ in general
grows in proportion to $N$, and therefore that ${\bm{\beta}} \cdot
{\bm{\beta}} / N {\cos}^{2}(\theta)$ is typically of the order of
unity for large $N$.  Thus the variance of the market-aligned
portfolio will be of the order of ${\bar{{\rho}^{2}}}_{mkt}$ for
large $N$, as would be expected. The variance of the
market-orthogonal portfolio, on the other hand, vanishes altogether
in proportion to ${N}^{-1}$ in the same limit of large $N$.  These
conclusions echo our results in \S 2, Eq.~(\ref{442}) et seq.

Note that the vanishing of the market risk for the market-orthogonal
portfolio, which is in addition to the vanishing of the ``diversifiable''
(or specific) risk expected for large $N$ (Elton and Gruber 1991), is a
specific result of leveraging coupled with hedging (or diversification).
Similarly, the infinite volatility and expected return levels of the $N-2$
remaining portfolios of this model underscore the dramatic levels of
volatility as well as return that can be expected of highly leveraged
portfolios.

We are now in a position to determine the composition of the
efficient frontier for the constant residual variance case. As
stated above, we find from the allocation rule of the efficient frontier that ${{X}^{crv}}_{\mu}=0
$ for $2 \leq \mu \leq N-1$, since the corresponding inverse
variances ${{Z}^{crv}}_{\mu}$ all vanish.  The three components of
the efficient frontier are the riskless asset together with ${{\bf
e}^{crv}}^{1}$ and ${{\bf e}^{crv}}^{N}$.  Furthermore, the latter
portfolio will be strongly disfavored relative to the former for
large $N$ since its variance grows in proportion to $N$ relative to
that of the former; cf. Eq.~(\ref{452}) et seq.  Indeed for
reasonably large $N$, the efficient frontier is essentially a
combination of ${{\bf e}^{crv}}^{1}$ and the riskless asset;
\begin{equation}
{X}^{crv}_{0} {\rightarrow} { {R}^{crv}_{1}-{\cal R} \over {R}^{crv}_{1}-
{R}_{0}}, \; {X}^{crv}_{1}{\rightarrow}{{\cal R}-{R}_{0} \over
{R}^{crv}_{1}-{R}_{0} }, \; \;{\rm as}\; {N \rightarrow \infty},
\label{4840}
\end{equation}
while
\begin{equation}
{X}^{crv}_{N}{\rightarrow} 0, \; {V}_{eff}=\vert {{\cal R}-{R}_{0} \over
{R}_{1}^{crv}-{R}_{0} } \vert {[{ {\bar{{\alpha}^{2}}} \over N {\sin}^{2}
(\theta)}]}^{1 \over 2} {\rightarrow}0 \;{\rm as}\; N \rightarrow \infty.
\label{485}
\end{equation}
This last property, i.e., the vanishing of the efficient portfolio
volatility (i.e., the market as well as the specific risk) in proportion
to ${N}^{-{1 \over 2}}$ in the limit as $N\rightarrow \infty$, also holds
for the general single-index model, as can be discerned from the results
of \S 2.  As discussed earlier, this total vanishing of the portfolio
volatility is a specific consequence of leveraging.

We close this section by summarizing the results established above.

{\bf Proposition 2.} {\it The principal portfolios of the constant
variance single-index model consist of a market-aligned portfolio ${{\bf
e}^{crv}}^{N}$, a minimum-volatility, market-orthogonal portfolio ${{\bf
e}^{crv}}^{1}$, and $N-2$ critically leveraged market-orthogonal portfolios
with infinite volatility and expected return, as given in Eqs.\
(\ref{446})-(\ref{452}).  Furthermore, as $N \rightarrow \infty$, the
efficient portfolio reduces to a combination of ${{\bf e}^{crv}}^{1}$ and
the riskless asset with a vanishing total volatility, as given in Eqs.\
(\ref{4840})-(\ref{485}).}

\section*{4. Concluding Remarks}

The main objective of this work, namely recasting the efficient
portfolio problem in terms of principal portfolios whereby the
selection is made from an uncorrelated set of portfolios instead
of the original asset set has been implemented in detail.   As we
have emphasized throughout, principal portfolios are the natural
instruments for analyzing the efficient frontier when short sales
are allowed.  More generally, they are the natural instruments for
any stock selection process based on the mean-variance
formulation. In effect, the analysis is transformed from the
original, correlated environment of individual assets to one of
uncorrelated portfolios, thus simplifying both the conceptual
framework as well as the practical procedures involved. This is of
course another example of the golden rule in applied analysis
which teaches us that when the basic object in the problem
involves a quadratic form, it is often advantageous to treat it in
the principal axes basis.

In order to illustrate the concepts and methods of this paper, we have
analyzed the single-index model as well as the constant residual variance
version of it in considerable detail.  Indeed our perturbative treatment
of the general single-index model for large asset sets has revealed
interesting new spectral features for that model.  In particular, the
bifurcation into market-aligned and market-orthogonal portfolios found in
\S 2 is an important observation on the volatility structure of the
model.  This is particularly so in view of the fact that for sufficiently
large asset sets the market-aligned portfolio is all but excluded from the
efficient frontier thereby eliminating the component of volatility
commonly referred to as market risk.  The constant residual variance
version of the model, while admittedly oversimplified, brings out the
above-mentioned bifurcation as well as the elimination of the market risk
in a clear and explicit manner.

The virtual elimination of efficient portfolio volatility for
reasonably large asset sets as well as the occurrence of large
volatility and expected return principal portfolios encountered in
our analysis demonstrate the combined effects of leveraging and
diversification.  As an example, consider the market-aligned
portfolio in the single-index model. This portfolio does not
involve any short positions and is unleveraged. Consequently, its
volatility is characteristic of the market volatility and does not
vanish for large asset sets.  The market-orthogonal portfolios, on
the other hand, are leveraged to varying degrees, a feature that
(together with hedging) largely immunizes them against the market
volatility.  These features are amplified and explicitly
demonstrated by the two distinguished principal portfolios of the
constant residual variance model.  The remaining portfolios of
this model, it may be recalled, are critically leveraged and as such
have infinite volatility and expected return.

We conclude by observing that the mean-variance description of risky asset
prices whereby short-term price variations are taken to be random
fluctuations has been a remarkably fruitful idea for describing the
dynamics of financial markets, its well known limitations notwithstanding.
We have attempted in this work to extend the utility of that idea by
providing a new analytical tool for its implementation.\\
\\

\noindent
{\bf REFERENCES}\\

\noindent
{Bachelier, L.} (1900) ``Th\'{e}orie de la Speculation,''
{\it Annals Scientifiques de l'Ecole Normale Superieure} \textbf{III-17}, 21-86.\\
{Black, Fischer and Scholes, Myron S.} (1973) ``The pricing of options
and corporate liabilities,'' {\it Journal of Political
Economy} \textbf{81}, 637-654.\\
{Elton, Edwin J. and Gruber, Martin J.} (1995) {\it Modern Portfolio
Theory and Investment Analysis}, 5th Ed., Wiley: New York.\\
{Fama, Eugene F. and Miller, Merton H.} (1972) {\it The Theory of Finance}, Holt Rinehart \& Winston.\\
{Ingersoll, J. E.} (1987) {\it Theory of Financial Decision Making}, Totowa, New Jersey: Rowman and Littlefield.\\
{Lintner, John} (1965) ``The valuation of risk assets and the
selection of risky in stock portfolios and capital budgets,'' {\it
Review of
Economics and Statistics} \textbf{47}, 13-37.\\
{Markowitz, Harry} (1952) ``Portfolio selection,'' {\it Journal of
Finance} \textbf{7}, 77-91.\\
{Markowitz, Harry} (1959) {\it Portfolio Selection: Efficient
Diversification of Investments}, New York: John Wiley \& Sons.\\
{Merton, Robert} (1972) ``An Analytic Derivation of the Efficient
Portfolio Frontier,'' {\it Journal of Finance and Quantitative
Analysis} \textbf{7}, 1851-1872.\\
{Mossin, Jan} (1966) ``Equilibrium in a capital asset market,'' {\it
Econometrica} \textbf{34}, 768-783.\\
{Ross, S. A.} (1976) ``The Arbitrage Theory of Capital Asset
Pricing,'' {\it Journal of Economic Theory} \textbf{13}, 341-360.\\
{Sharpe, William F.} (1963) ``A simplified model for portfolio
analysis,'' {\it Management Science} \textbf{11}, 277-293.\\
{Sharpe, William F.} (1964) ``Capital asset prices: A theory of
market equilibrium under conditions of risk,'' {\it Journal of
Finance} \textbf{19}, 425-442.\\
{Tobin, James} ( 1958) ``Liquidity preference as behavior towards
risk,'' {\it Review of Economic Studies} \textbf{25}, 65-86.\\
{Partovi, M. Hossein and Caputo, Michael R.} (2004), ``Principal Portfolios: Recasting the Efficient Frontier,'' \textit{Economics Bulletin} \textbf{7}, 1−10.\\
{Poddig, Thorsten and Unger, Albina} (2012), ``On the robustness of risk-based asset allocations'', \textit{Financial Markets and Portfolio Management} \textbf{26}, 369-401.\\
{Kind, Christoph} (2013) ``Risk-Based Allocation of Principal Portfolios,'' SSRN: http://ssrn.com/abstract=2240842, 2013.\\


\end{document}